\documentclass{ws-procs9x6}

\begin{document}

%\thispagestyle{empty}
%\begin{figure}[htb]
%  \centering
%  \includegraphics[width=0.99\textwidth,angle=0]{cr06_018.eps}
%\end{figure}
%\clearpage\newpage

\title{\vspace*{-13mm} Getting ready for Physics at the LHC\\ with the CMS Detector}

\author{\vspace*{-5mm} V.~Drollinger\footnote{\uppercase{W}ork supported in part by the \uppercase{E}uropean \uppercase{C}ommunity's \uppercase{H}uman \uppercase{P}otential \uppercase{P}rogramme under contract \uppercase{HPRN}-\uppercase{CT}-2002-00326, [\uppercase{V}.\uppercase{D}.].}}

\address{Universit\a`a di Padova, \\
Dipartimento di Fisica "Galileo Galilei", \\ 
Via F. Marzolo, 8, 35131 Padova, Italy\\ 
E-mail: drollinger@pd.infn.it}

\maketitle

\abstracts{\vspace*{-3mm}
In order to get ready for physics at the LHC, the CMS experiment has to be set up for data
taking. The data have to be well understood before new physics can be investigated.
On the other hand, there are standard processes, well known from previous experiments
and from simulation, which will help to understand the data of the detector in the
early days of the LHC.}

%%%%%%%%%%%%%%%%%%%%%%%%%%%%%%%%%%%%%%%%%%%%%%%%%%%%%%%%%%%%%%%%%%%%%%%%%%%%%%%%%%%
\vspace*{-8mm}
\section{Goals and Needs}

The main goals for physics at the LHC are searches for new physics and precision measurements. One of the main tasks at the LHC is to probe the existence of the Higgs boson or the source of electroweak symmetry breaking in general. Furthermore, many models with new physics wait to be investigated experimentally. In particular, for phenomena that are expected to show up at high energies, the LHC is reaching completely new territory. In general, precision measurements are somewhat less important than searches at hadron colliders, but there are cases where high statistics is a benefit, or new production channels are open. A prominent example is precision physics with top quarks.

Knowing the goals it is not difficult to imagine what the needs are.\linebreak First of all, it is essential to have a well working detector, which allows continuous data taking over several hours. The quality of these data has to be high in the sense that none of the detector subsystems has major failures, and that the resolutions are sufficient to carry out the triggering, reconstruction and selection of events with a high precision. In addition to a high event-reconstruction efficiency, it is necessary to have a good particle identification in order to keep the corresponding fake rates as low as possible, which is essential to avoid the huge backgrounds at hadron colliders.  Furthermore, the detector needs to be synchronized, aligned and calibrated in order to reach the ultimate detector performance. In particular at the LHC, a high performance trigger system is needed in order to reduce the event rate by about six orders of magnitude.

Since it is clearly impossible to review this complex subject in five pages, a brief overview is presented and a few examples are given.

%%%%%%%%%%%%%%%%%%%%%%%%%%%%%%%%%%%%%%%%%%%%%%%%%%%%%%%%%%%%%%%%%%%%%%%%%%%%%%%%%%%
\section{The CMS Experiment}

\begin{figure}[htb]
  \centering
  \includegraphics[width=0.62\textwidth,angle=0]{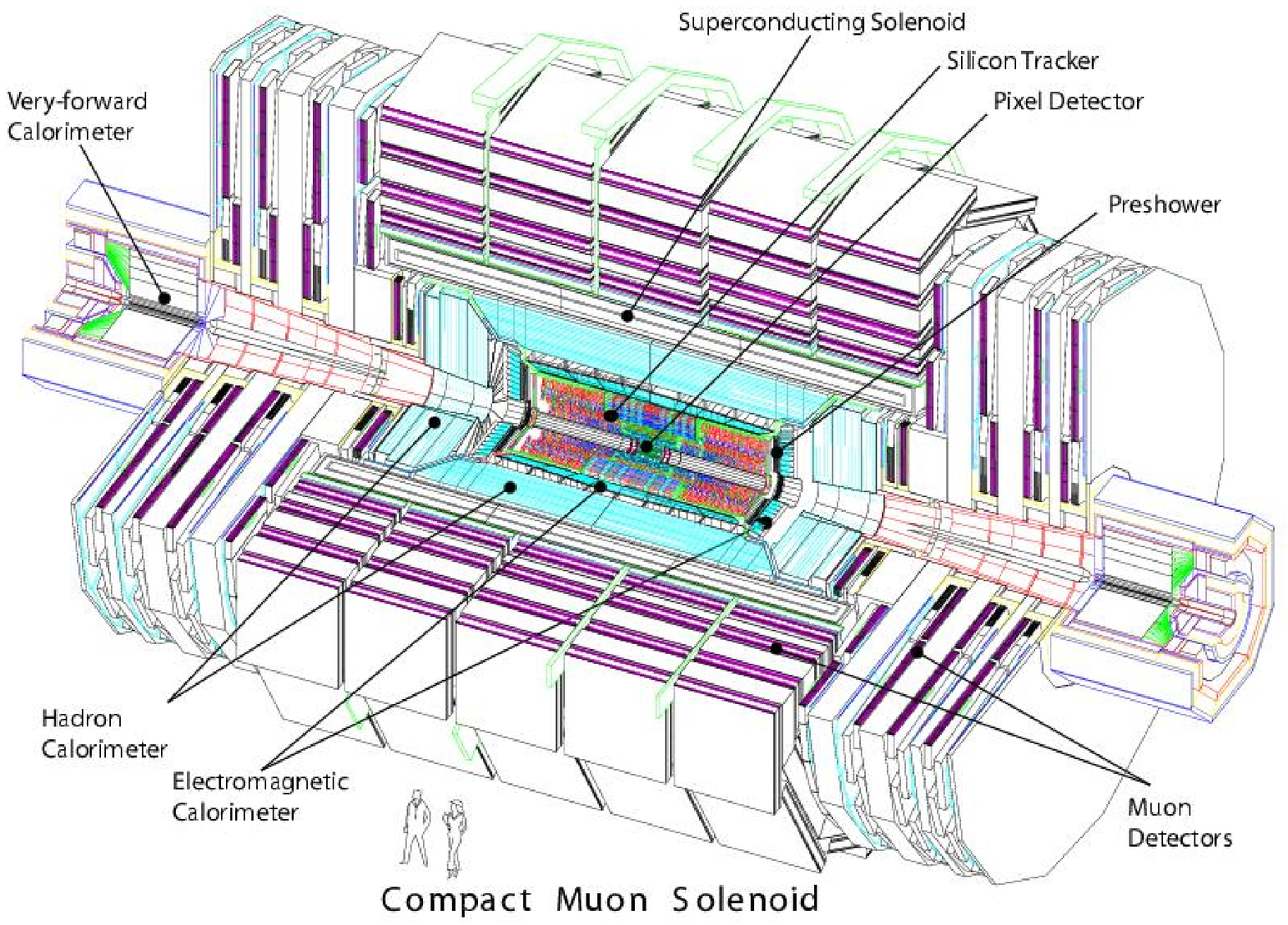}
  \hspace*{6mm}
  \includegraphics[width=0.45\textwidth,angle=90]{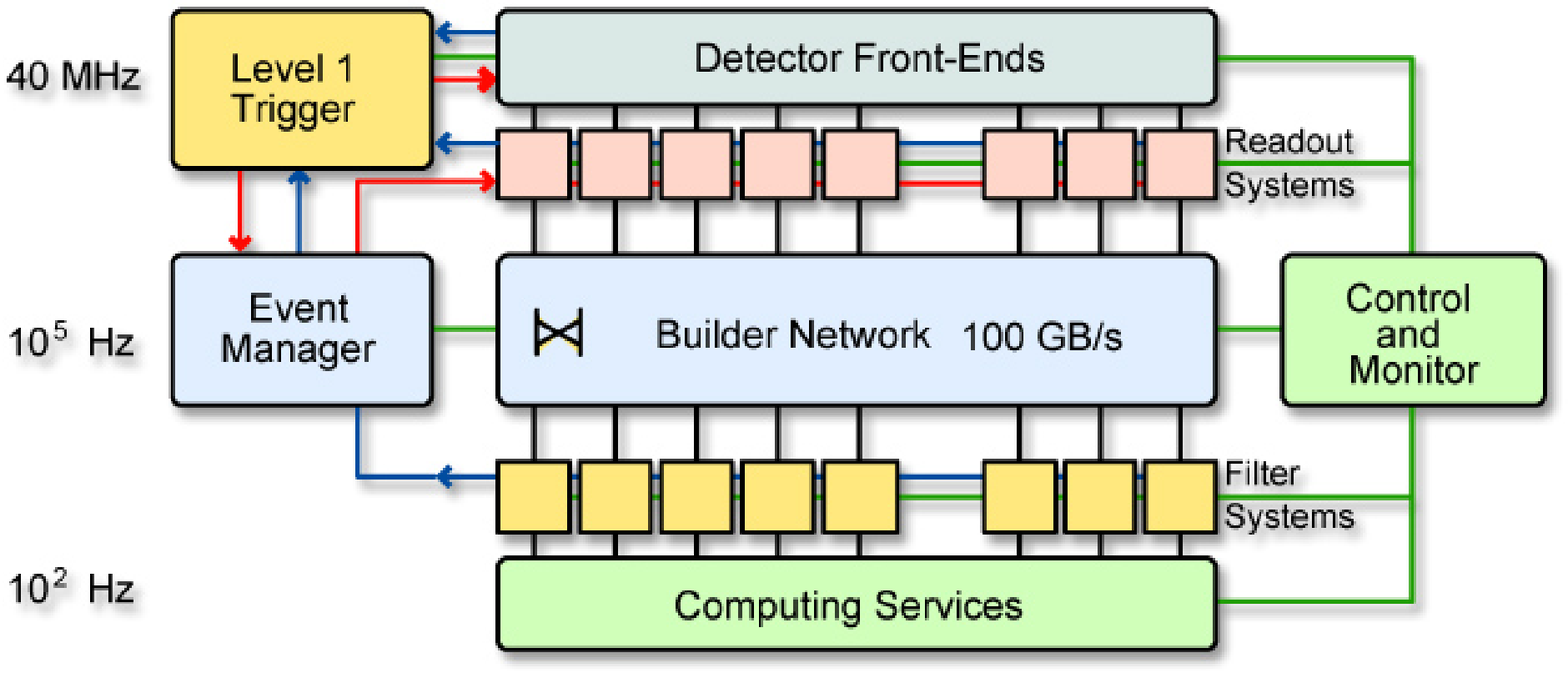}
  \vspace*{-4mm}
  \caption{Drawing of the CMS detector, and of the trigger and DAQ scheme.}
  \label{fig:vcms}
\vspace*{-1mm}
\end{figure}

\begin{figure}[htb]
  \centering
  \includegraphics[width=0.66\textwidth,angle=0]{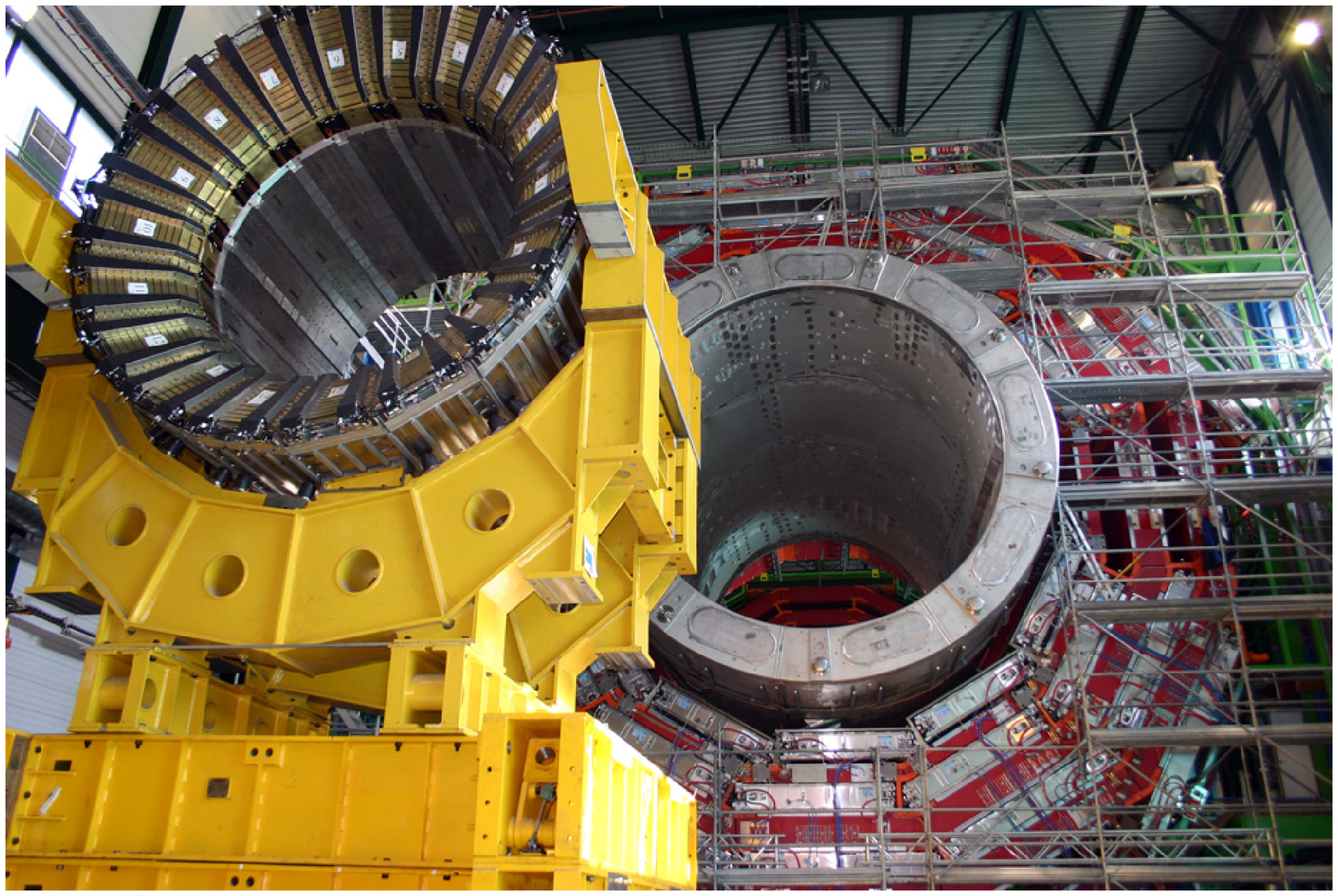}
  \includegraphics[width=0.33\textwidth,angle=0]{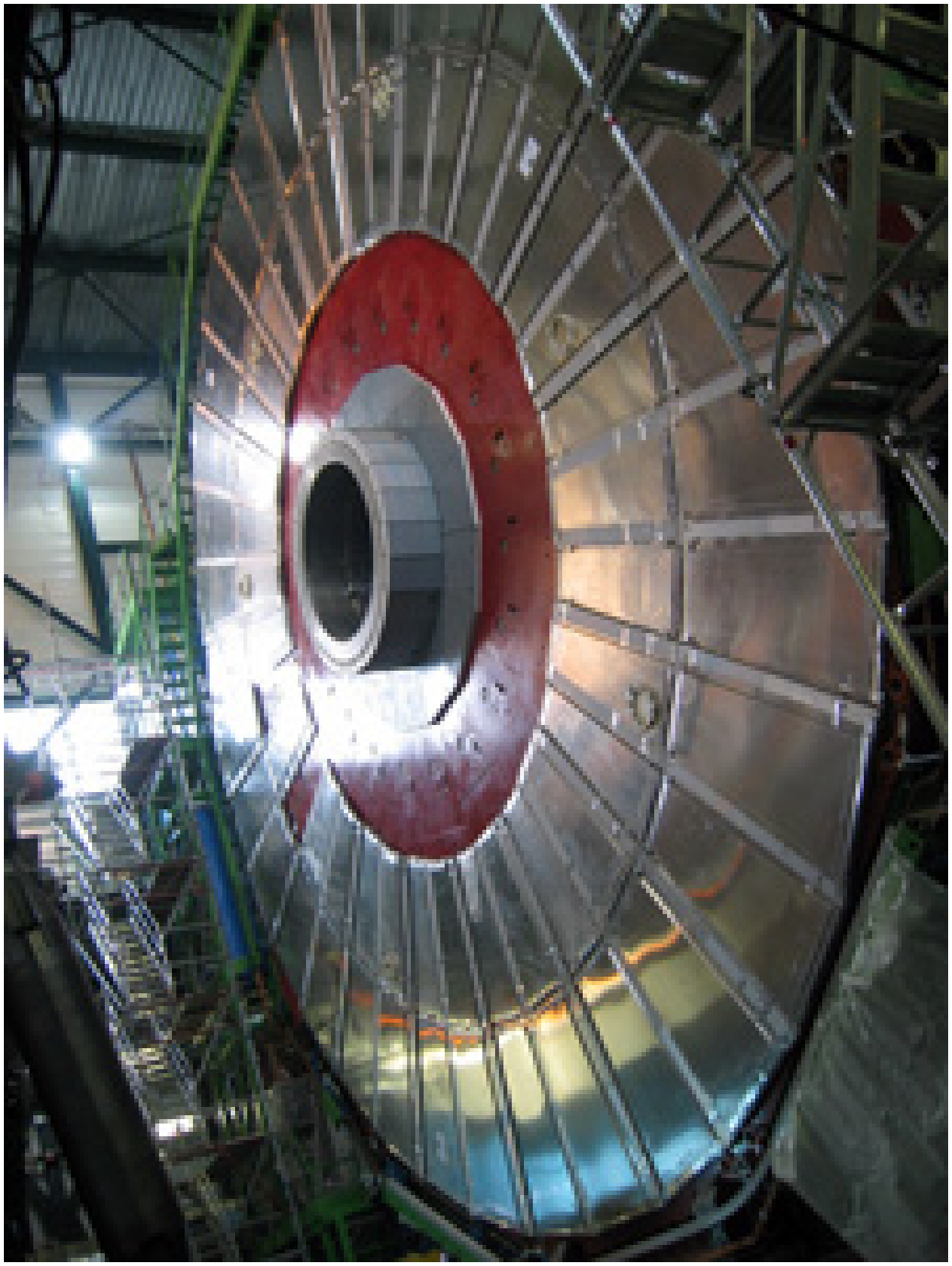}
\vspace*{-6mm}
  \caption{Left: barrel components (front: hadron calorimeter, back: magnet and muon system) of the CMS detector. Right: forward components (muon system) of CMS.}
  \label{fig:cms}
\vspace*{-2mm}
\end{figure}

The CMS detector\cite{ref:cms} is a general purpose detector, which is designed to meet the goals mentioned above. Figure~\ref{fig:vcms} shows schematic views of the CMS detector and the structure of the trigger and Data Acquisition (DAQ).
\vspace*{-2mm}
The CMS detector is designed like a classical collider detector with tracker (including pixel vertex detectors) and muon system acceptance up to $\eta$~=~2.5, an electromagnetic calorimeter acceptance up to $\eta$~=~3, and hadron calorimetry up to $\eta$~=~5. The calorimeters are designed to be able to contain jets in the TeV range, and a superconducting magnet provides a magnetic field of 4~T, which allows transverse momentum measurements up to about 1~TeV. A large fraction of the CMS detectors has been built and is currently tested, as illustrated in Figs.~\ref{fig:cms} and~\ref{fig:cosmic}.

The trigger and DAQ system is designed to cope with initial event rates of 40~MHz (100~Tbyte/s), which is reduced by the Level~1 (L1) trigger to 100~kHz (100~Gbyte/s), and further reduced to about 100 Hz (100~Mbyte/s) by the High Level Trigger (HLT). Whereas the L1 trigger consists of dedicated trigger boards, the HLT trigger is running CMS reconstruction and selection software on a computer farm.

Also on the software, simulation and reconstruction side a lot of progress have been made and is described in detail in Ref.~\cite{ref:cms}. The preparation for physics analyses is going on and will be documented in an upcoming CMS Physics Technical Design Report. It is also important to setup strategies on how to perform the analyses, such as background normalization from the data itself, employing advanced analysis techniques.

%%%%%%%%%%%%%%%%%%%%%%%%%%%%%%%%%%%%%%%%%%%%%%%%%%%%%%%%%%%%%%%%%%%%%%%%%%%%%%%%%%%
\section{From Cosmic Rays to first LHC Collisions}

\begin{figure}[htb]
  \centering
  \includegraphics[width=0.49\textwidth,angle=0]{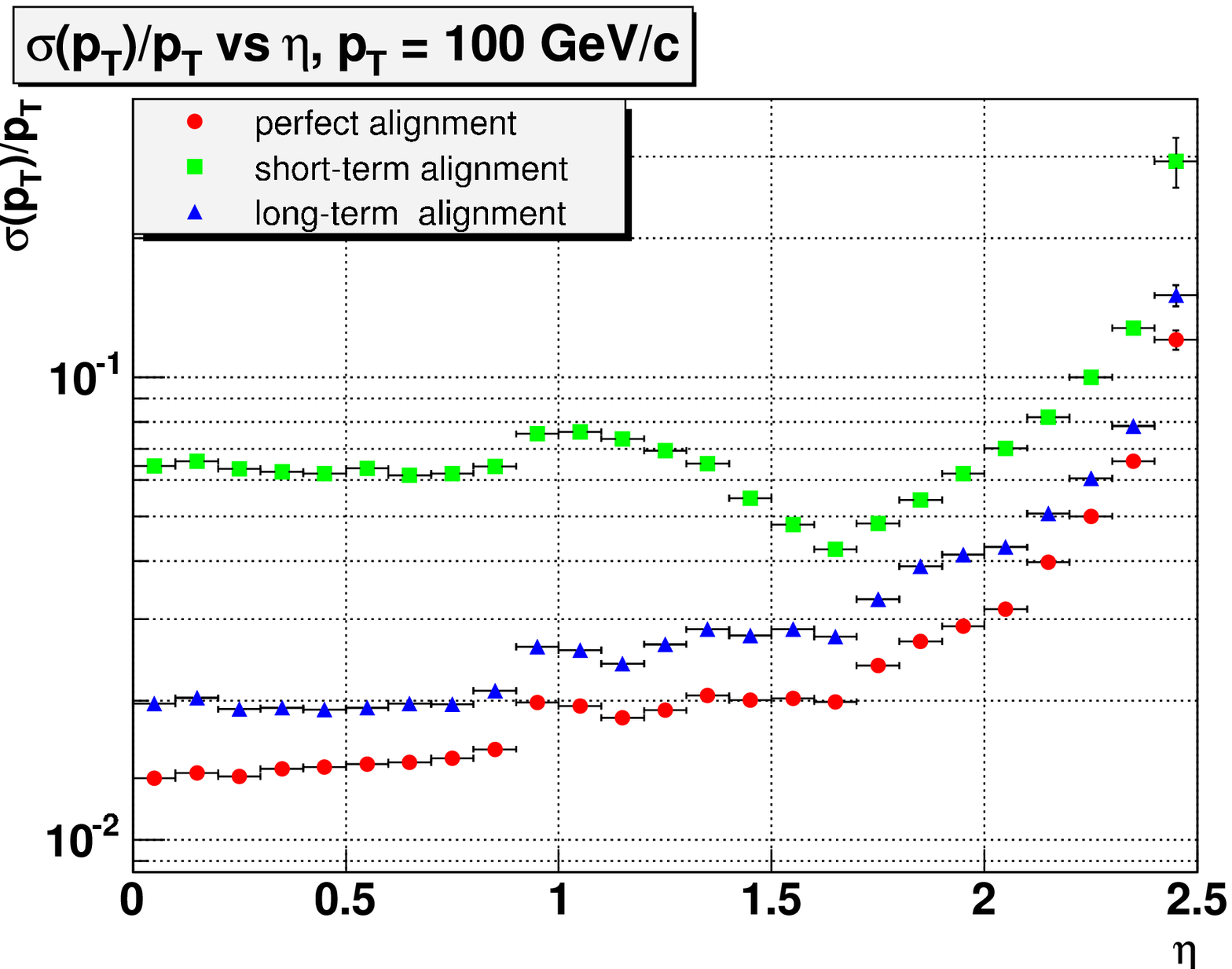}
  \includegraphics[width=0.49\textwidth,angle=0]{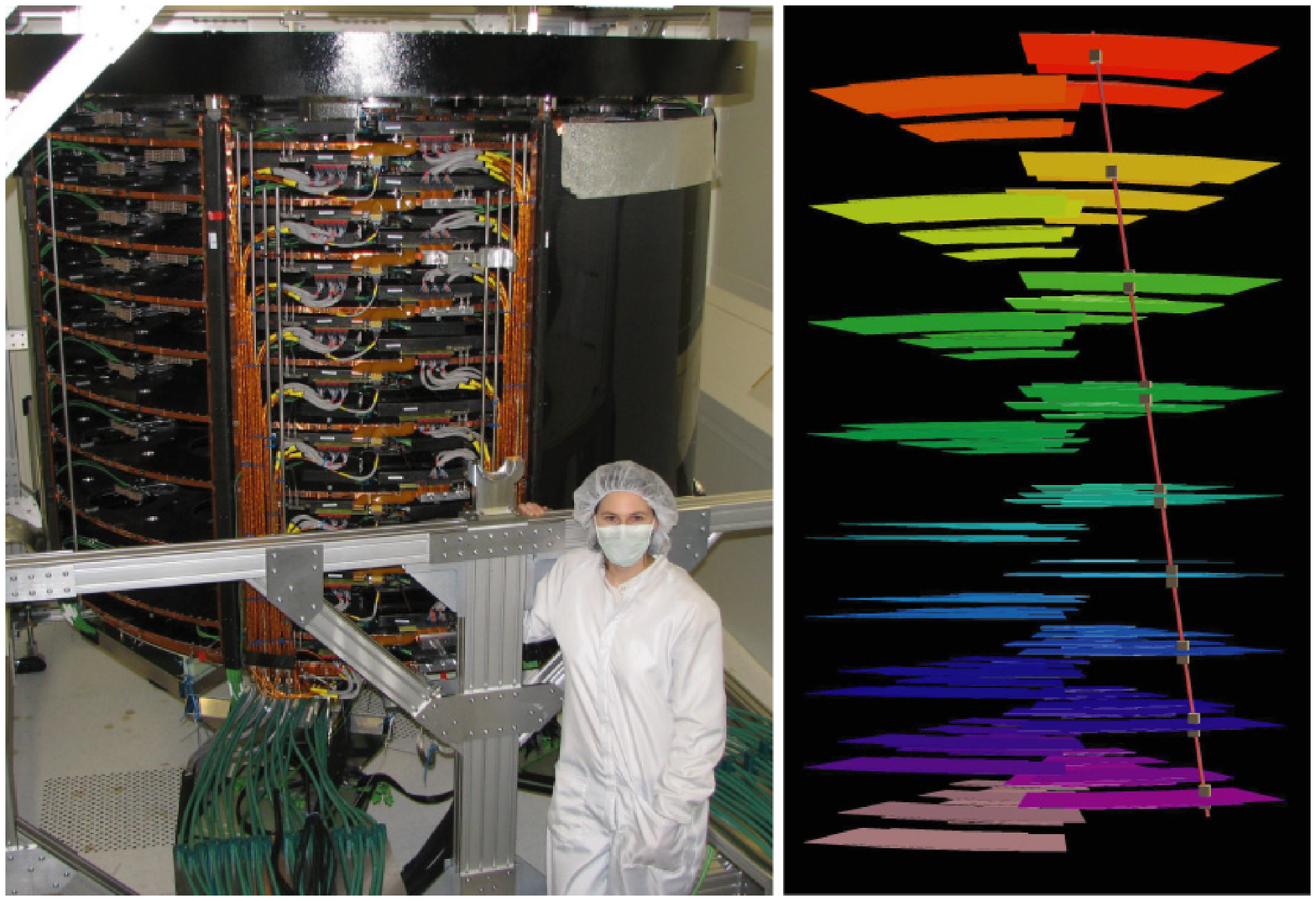}
  \caption{Left: simulation of three tracker alignment scenarios with single muons. Right: cosmic rays seen with one endcap of the inner tracker of CMS.}
  \label{fig:cosmic}
\vspace*{-2mm}
\end{figure}

After many test beams for individual detector components has been carried out, the main focus at present is the test of groups of detectors with cosmic rays. One example of the inner tracker is shown in Fig.~\ref{fig:cosmic}. A large scale test, including many different detector types and the entire magnet with all iron return yokes, the ``Magnet Test Cosmic Challenge'' in preparation.

In order to arrive at a fully operational experiment, many steps are necessary and the comparison of simulation and data is very helpful to understand each step. On one hand, the simulation is employed to model the detector and to make accurate predictions of many things like trigger rates, efficiencies, resolutions, fake rates. On the other hand, the data can be used to tune the simulation to describe the reality even more accurately.
One example of a detailed simulation is the tracker alignment, shown in Fig.~\ref{fig:cosmic}, where the transverse momentum momentum resolution of the CMS tracker is predicted for three alignment scenarios. The performance of the long-term alignment scenario gets close, but is not identical, to the performance of a perfectly aligned tracker.

The LHC pilot run in the year 2007 will be the first source of collision data. The numbers of events of important Standard Model processes, expected for one particular scenario\cite{ref:gigi}, are summarized in Table~\ref{tab:pilot}.
Even for a low luminosity and a short running period the number of minimum bias events and di-jet events is huge. These events are important to study the calorimeter response and to understand the general structure, e.g. the underlying event, of events at the LHC. Energetic leptons, coming from W or Z decays, are produced with sufficient statistics in order to study electron and muon triggering, reconstruction, identification and the corresponding efficiencies. Only a moderate number of top quarks, important as a signal and as a background, is expected in the beginning of the LHC.
\vspace*{-1mm}
\begin{table}[hbt]
\tbl{Expected number of events for the pilot run of the LHC with ${\rm \sqrt{s_{pp}} =}$~14~TeV. One month of running with a luminosity of $L = $ 10$^{30}$ cm$^{-2}$ s$^{-1}$ is assumed.}
{\footnotesize
\begin{tabular}{@{}|c|ccccc|@{}}
\hline
{} &{} &{} &{} &{} &{}\\[-1.5ex]
  process & min.bias  & jets$_{\, E_T > 60\ {\rm GeV}}$ & ${\rm W^\pm \to \ell^\pm \nu}$ & ${\rm Z \to \ell^+ \ell^-}$ & ${\rm t\bar{t} \to \ell^\pm \nu\, jets}$\\[1ex]
\hline
{} &{} &{} &{}  &{}&{}\\[-1.5ex]
\# events &  $\gtrsim$ 10$^{12}$ &   $\gtrsim$ 10$^{8}$ &  $\gtrsim$ 5 $\times$ 10$^{4}$ &  $\approx$ 5 $\times$ 10$^{3}$ & $\approx$ 3 $\times$ 10$^{2}$ \\[1ex]
\hline
\end{tabular}\label{tab:pilot}}
\vspace*{-13pt}
\end{table}
\vspace*{1mm}
Even during the pilot run, new physics could be potentially produced, but it is rather unlikely that such events can be recognized without understanding the detector in detail.

%%%%%%%%%%%%%%%%%%%%%%%%%%%%%%%%%%%%%%%%%%%%%%%%%%%%%%%%%%%%%%%%%%%%%%%%%%%%%%%%%%%
\section{Help from Others}
Many aspects, discussed above, are not completely new. Clearly, a lot of experience on experimental aspects is available from other (collider) experiments, and an effort has been started to propagate the knowhow to the LHC experiments\cite{ref:otherex}. In addition, it is very important have a good idea of what kind of physics to expect at the LHC and collaborations with theorists have been established\cite{ref:theory}. Theoretical predictions are required for setting up physics analyses, but also for designing the detector and optimizing the reconstruction software.

\begin{figure}[htb]
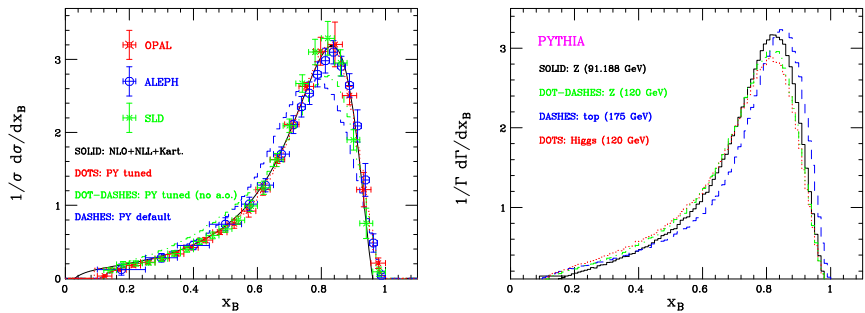

  \centering
  \includegraphics[width=0.48\textwidth,angle=0]{bfrag_py_ee.eps}
  \hspace*{2mm}
  \includegraphics[width=0.48\textwidth,angle=0]{bfrag_zbhp.eps}
  \vspace*{-6mm}
  \caption{Bottom fragmentation in ${\rm Z \rightarrow b\bar{b}}$ decays of PYTHIA tuned to ${\rm e^+ e^-}$ data (left) and predictions for ${\rm t \rightarrow W b}$ and ${\rm h \rightarrow b \bar{b}}$ decays (right).}
  \label{fig:bfrag}
\end{figure}

One concrete example of how the LHC experiments can benefit from previous collider experiments and from theoretical work is bottom fragmentation in top (${\rm t \rightarrow W b}$) and Higgs (${\rm h \rightarrow b \bar{b}}$) decays\cite{ref:bfrag}. In order to understand bottom fragmentation in these processes, a Monte Carlo event generator is tuned to ${\rm e^+ e^-}$ data by matching the B-hadron spectra $x_B$, the energy fraction of the B-hadron normalized to the energy of the b-quark, of the ${\rm Z \rightarrow b \bar{b}}$ decays, shown in Fig.~\ref{fig:bfrag}.
\vspace*{-1mm}
\vspace*{-4mm}
In a second step, the tuned event generator is employed to predict $x_B$ spectra for other processes than ${\rm Z \rightarrow b \bar{b}}$, namely ${\rm t \rightarrow W b}$ and ${\rm h \rightarrow b \bar{b}}$ which are of interest at the LHC and at the TeVatron, too.

%%%%%%%%%%%%%%%%%%%%%%%%%%%%%%%%%%%%%%%%%%%%%%%%%%%%%%%%%%%%%%%%%%%%%%%%%%%%%%%%%%%
\section{Conclusions}

The main goal, studying physics at the LHC, is coming closer: a big fraction of the CMS detector has been built and is currently tested with cosmic rays. In context with the CMS Physics Technical Design Reports, many aspects of the detector an physics performance have been studied in detail in preparation for the CMS physics program.

%%%%%%%%%%%%%%%%%%%%%%%%%%%%%%%%%%%%%%%%%%%%%%%%%%%%%%%%%%%%%%%%%%%%%%%%%%%%%%%%%%%

%

\begin{thebibliography}{0}

\bibitem{ref:cms} CMS Collaboration, {\tt cmsinfo.cern.ch}, {\tt cmsdoc.cern.ch/cms/cpt/tdr},\\
                  and references to Technical Design Reports therein.

\bibitem{ref:gigi} G.~Rolandi, private communications.

\bibitem{ref:otherex} TeV4LHC, {\tt conferences.fnal.gov/tev4lhc};\\
                      HERA/LHC, {\tt www.desy.de/\%7Eheralhc}.

\bibitem{ref:theory} Physics at TeV Colliders, {\tt wwwlapp.in2p3.fr/conferences/LesHouches}.

\bibitem{ref:bfrag} G.~Corcella and V.~Drollinger, {\it Nucl. Phys.} {\bf B730}, 82-102 (2005).

\end{thebibliography}
\end{document}